*COTS MOS Dosimetry on the MeMOSat Board, Results After 2.5 Years in Orbit*

José Lipovetzky (1*, 2,3,4), Mariano Garcia-Inza (1,3), Macarena Rodríguez Cañete (1), Gabriel Redin (1), Sebastián Carbonetto (1,3), Martín Echarri (5*), Federico Golmar (3,5,6), Fernando Gomez Marlasca (2), Mariano Barella (2,3,5), Gabriel Sanca (6), Pablo Levy (2,3,6), Adrián Faigón (1,3)

(1) Facultad de Ingeniería, Universidad de Buenos Aires || Paseo Colón 850, Ciudad de Buenos Aires, Argentina || Phone +541152850819

(2) Comisión Nacional de Energía Atómica || Bustillo 9500, Centro Atómico Bariloche, Bariloche, Argentina

Phone +54 294 444 5100 x 5349

(3) Consejo Nacional de Investigaciones Científicas y Técnicas || (4) Instituto Balseiro, Argentina

Bustillo 9500, Centro Atómico Bariloche, Bariloche, Argentina

(5) Centro de Micro y Nanoelectrónica del Bicentenario - Instituto Nacional de Tecnología Industrial (INTI), Argentina

(6) Escuela de Ciencia y Tecnología, Universidad de San Martín (UNSAM), Argentina

Mails: jose.lipovetzky@ieee.org , fgolmar@inti.gob.ar , magarcia@fi.uba.ar, afaigon@fi.uba.ar

*formerly

**Abstract**: We present the results after 2.5 years in orbit of Total Ionizing Dose (TID) measurements done using Metal Oxide Semiconductor (MOS) dosimeters on the MeMOSat board. The MeMosat board was launched on July 19th 2014 at the BugSat-1 "Tita" microsatellite developed by Satellogic to stay at LEO. We used as dosimeters p-channel Commercial Off The Shelf (COTS) MOS transistors with gate oxides of 250 nm. Before launch, a subset of transistors with similar drain current to voltage (I-V) curves where selected from a group of 100 devices. The temperature dependence of the (I-V) curves was studied to find the minimum temperature coefficient biasing point. Then, a calibration subgroup of sensors was irradiated using a $^{60}$Co gamma source to study their response to TID, showing responsivities of ~75 mV/krad when the sensors are irradiated without gate bias. Also, the post irradiation response of the sensors was monitored, in order to include a correction for low dose rate irradiations, yielding 30 mV/krad. A biasing and reading circuit was developed in order to allow the reading of up to 4 sensors. The threshold voltage was monitored during different periods of the mission. After 2.5 years in orbit, the threshold voltage of the sensor mounted on the MeMOSat Board had a $V_T$ shift of approximately 35 mV corresponds to a dose of 1.2 krads.

## I. Introduction

Metal Oxide Semiconductor (MOS) dosimeters are MOS transistors which allow the quantification of the Total Ionization Dose (TID) through the shift of the Threshold Voltage ($V_T$) caused by buildup of positive charge in the oxide and the generation of interface traps [1,2]. MOS dosimeters have been used in space applications [3-6] and medical applications [7,8]. To allow high sensitivities, MOS dosimeters are usually manufactured in ad-hoc processes with gate oxides of hundreds of nanometers or even few micrometers, thicker than regulars MOS transistors [1-4,7,9]. Since the responsivity to TID is approximately proportional to the oxide thickness $t_{ox}$ [10,11].

However, several works have dealt with the use of Commercial Off The Shelf MOS transistors as MOS dosimeters in medical applications [12] or industrial applications [13,14]. This work proposes the use of a COTS transistor as a low cost sensor for on-board dosimetry in a Cubesat.

## II. Devices and initial calibration

The devices used in this work as sensors are COTS p-channel MOS enhancement transistors with a gate oxide thickness of ~250nm. The oxide thickness---not informed by the manufacturer---was estimated using gate tunnel current vs gate oxide voltage characteristics of the sensors during Fowler Nordheim current injection [13]. The MOS devices are packaged in a TO-72 metal case.

It is known that the responsivity to TID of different MOS dosimeters can have a high dispersion—up to 30%---if are manufactured in different Si wafer [13-14]. This dispersion would introduce errors in the quantification of dose if the devices are not individually calibrated. Thus, initially, a subset of 15 devices from a batch of 100 MOS transistors where selected using as criterion to have similar drain current to gate voltage (I-V) characteristics, i.e. similar $V_T$ and transconductance values. These devices where calibrated.

### A. Temperature Dependence
The temperature dependence of the I-V curve of the devices selected for calibration was studied. The I-V curve of MOS dosimeters is affected by two main temperature dependencies, the decrease of $V_T$ and the decrease of the transconductance with temperature. Usually, both effects compensate for a given drain current value, known as Zero Temperature Coefficient (ZTC) current ($I_{ZTC}$). Usually in MOS dosimetry, this point of the I-V curve is used to measure the $V_T$ shift [4]. Fig.1 presents the I-V curves measured at four different temperatures from 0ºC to 60ºC, a temperature wider than the temperature fluctuations expected during the mission. It can be observed that the ZTC point of the curve is at $I_{ZTC}$=270 μA. However, due to circuital restrictions, the dosimeters where finally biased with a 200 μA drain current to quantify the $V_T$ shift on the board. This introduced a small temperature dependence of the reading as will be seen later.

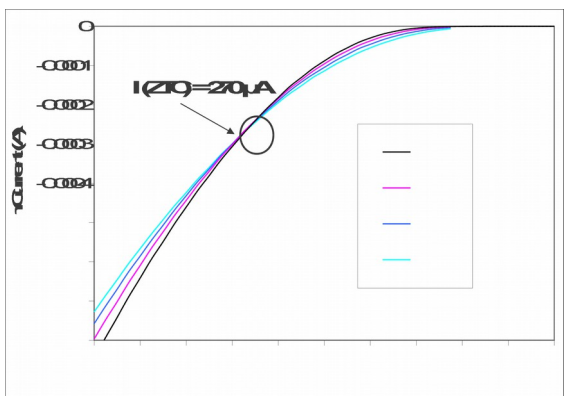

Fig. 1. I-V characteristics of the selected devices.

### B. TID Response
The response of the sensors to TID was studied to obtain the responsivity of the sensors. It is known that the responsivity of a MOS dosimeter depends on the gate voltage sustained during the irradiation. Usually, a positive gate voltage will increase the responsivity. However, a constraint imposed by the satellite platform to the board with the dosimeters was that the board would not be powered during most of the time of the mission. This implies that the sensors would be biased during most of the mission time with zero volts between all device terminals.

The response to irradiation of the sensors was done using a $^{60}$Co gamma rays source, at a dose rate of ~0.5Gy/min (50rads/min). Some sensors where biased with a gate voltage of zero volts (as would be on the satellite), and others with a gate voltage of 9V. Fig. 2 presents the $V_T$ shift as a function of dose during irradiation up to 1.6 and 1.8 Gy with gate voltages of 9 V and 0 V respectively. The responsivities of the devices where 58mV/Gy and 6.6mV/Gy for 9 V and 0 V bias. Figure 3 shows the shift of the I-V curve after irradiation in a device, showing how a shift in the gate voltage required to apply a given drain current can be used as a dosimetric magnitude.

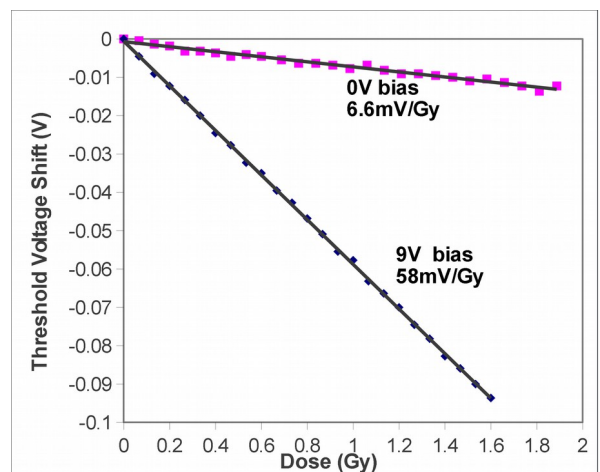

Fig. 2. Response to TID during $^{60}$Co irradiation of one sample of MOS dosimeters with 0V and 9V gate bias.

The experiment of Fig. 2 was repeated with an incremental dose of 1.5Gy with zero volts bias to the 15 devices of the set, yielding responsivities from 6.6mV/Gy to 7.5mV/Gy, proving that each device needs an individual calibration despite of having similar initial $V_T$ values. This pre-use irradiation allows the initial calibration of each device.

## C. Long Term Annealing.

A difficulty with the use of MOS dosimeters is fading, i.e. the long term recovery of $V_T$ with time, a problem associated to the neutralization of oxide trapped charge via tunneling or thermal excitation of electrons [1-4]. Annealing can introduce errors in the quantification of TID if the sensors are calibrated at a high dose rate and then the dosimetry is carried out at such a low dose rate that annealing causes a significant recovery in $V_T$, underestimating the absorbed dose. This feature is also known as apparent dose rate dependency of the response of the sensor [4]. It has been reported that the final $V_T$ shift after a short irradiation at a high dose rate followed by long term annealing is approximately similar to the $V_T$ shift obtained after the exposure to the same dose in the same annealing time at a much lower dose rate. To investigate into this, one of the calibrated devices was irradiated with the ⁶⁰Co source up to 20.3 Gy using a gate bias voltage of 0V before the launch of the satellite. The device was kept on Earth during the mission time with all terminals grounded, to evaluate the recovery of $V_T$ after the mission time. Figure 3 shows the I-V curves of the fresh MOS sensor, after irradiation and after 2.5 years of annealing. Initially, $V_T$ shifts -152 mV. But then, after anneal, $V_T$ recovers 92 mV, a 60%. Thus, the final after anneal $V_T$ shift happens to be 60 mV, meaning that the real low dose rate sensitivity of the dosimeter is only 3.0 mV/Gy. This is an important result, since the irradiation in space will be performed in a scale time of years instead of minutes.

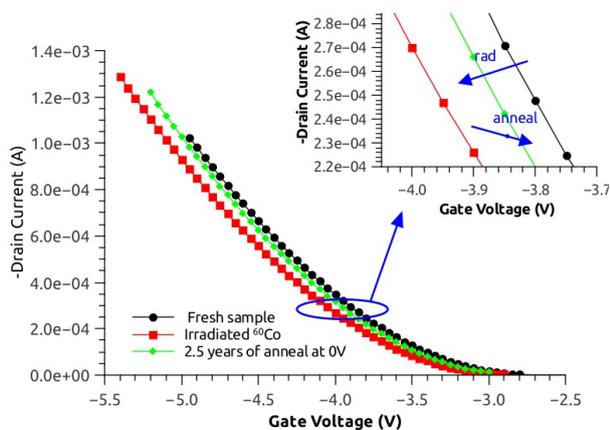

Fig. 3, change in I-V curves after a 20.3Gy (2.03 krads) irradiation with 0 V gate bias and the recovery after 2.5 years.

## III. Memosat Board and Reading Circuit

Two COTS MOS transistors were used as sensor was mounted on the MeMOSat-01 board, which was part of the payload of the BugSat-1, a microsatellite developed by Argentine company Satellogic [15], launched from the Dombarovsky air base (Russia) in June 19th, 2014. The BugSat-1 is on a LEO orbit with 620 km of altitude and an inclination of 97.9 degrees and weights 22 kg.

MeMOSat-xx is a reconfigurable platform for testing ReRAM nonvolatile memories in the satellite environment. The first launched board, MeMOSat-01 was designed to perform target specific tests on two ReRAM HfO₂, generating reports including experimental and system parameters, reported periodically to Earth via interaction with the satellite. MeMOSat-xx was the first step of LabOSat , a comprehensive project to perform experiments in space. The MeMOSat-01 board is mounted close to the center of the Bugsat-1 satellite, surrounded by a large amount of mass in most directions.

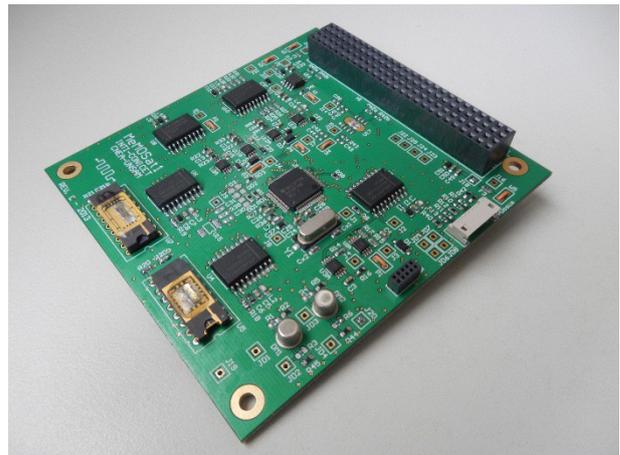

Fig. 4. The MemoSat board before mounting the MOS dosimeter.
The sensors are mounted on the bottom corner of the board.

The circuit designed to read $V_T$ shifts of the sensor was designed taking in account that the platform of the satellite would not provide power during most of the time. Moreover, since the main objective of the mission is to test the satellite itself, the MeMOSat board would be turned on only during short intervals, during specific periods of time. Thus, the circuit had to ensure that the sensor would have all the time the zero volts gate bias, except for very short reading times in which the reading drain current ($I_{READ}$) would be applied to measure the $V_T$ shift. Also, to be able to measure in future missions the dose in different points of the satellite, the circuit should allow the reading of four devices using the same analog to digital converter (ADC).

The schematic design of the circuit used in the board is presented in Fig. 5. When the analog switches S10 and S20 are open, the sensor (QRAD-FET) is in the "Exposure" mode, with all terminals are grounded through 1 MΩ resistors. When the switches are closed, the operational amplifier U2 applies on the gate of the sensor the gate voltage required to have a fixed drain bias current of $I_{READ}$(9V-Vref)/Rref. The gate voltage was reduced using a resistive divider and read using a 12-bit ADC, yielding a resolution of 2.4mV per ADC count. The analog switches are in fact the switches of an analog multiplexer, allowing the sequential reading of four devices with the same electronics.

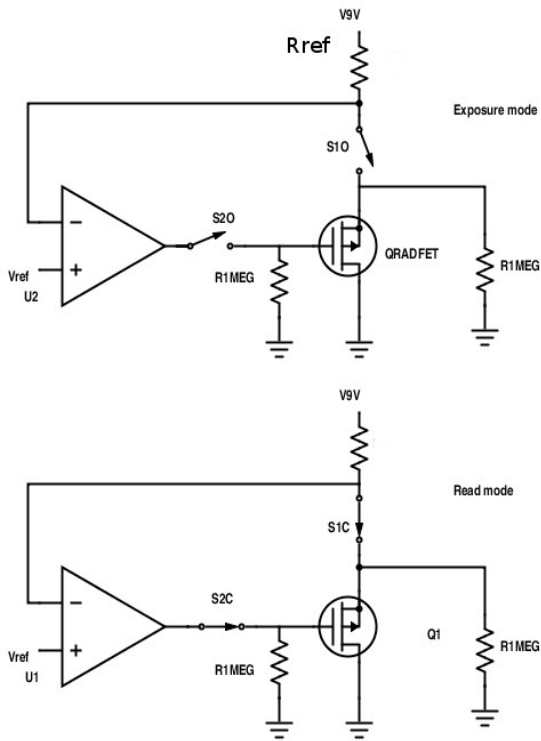

Fig 5. Bias and reading circuit.

Device restrictions forced the use of $I_{READ} = 200\mu A$, which was, as explained before, not exactly the $I_{ZTC}$ value. Thus, the temperature dependence of the voltage reading was estimated from I-V curves at different temperatures, yielding ~296μV/°C.

## IV. Results and Discussion

The sensors were mounted on the MeMOSat board, included on the satellite and launched. Before launch, the initial $V_T$ value was read. The satellite began to transmit dosimetric information after all the initial checks and tests were performed on the rest of the platform, 164 days after launch. Only one of the two sensors resulted to be functional after this period.

Figure 6 presents the $V_T$ values obtained during the mission. There are two periods of time in which the information was not recorded, at the beginning of the mission and during the second year.

It can be observed that $V_T$ has a clear tendency to reduce, with a high dispersion. This dispersion---of ~±2ADC counts or ~±5mV---is attributed to thermal variations in the satellite, causing an uncertainty in the reading of $V_T$ due to the fact that the read current was not set exactly to the ZTC point of the I-V curve of the device. This dispersion is consistent with a temperature amplitude of 32°C, which agrees with thermal information provided by the satellite manufacturer which reported less than 40°C of temperature amplitude on the board. Unfortunately, we did not have available measurements of temperature done at the same times of $V_T$ readings. New editions of the board include on board temperature measurements to allow thermal correction of dosimetric magnitudes.

The threshold voltage shift recorded during the mission is approximately -35mV. Taking in account that the calibrated sensitivity of the sensor at low dose rates was 3.0mV/Gy, the dose measured by the detector is ~12 Gy---i.e.1.2 krads.

This dose is surprisingly lower than what is usually expected in a LEO orbit, higher than tens of Gy/year [16] in most satellites. However, the explanation for the low dose measured on the experiment is that since the board is placed closed to the center of the satellite, most of the mass of the rest of the system serves as a shielding for most particles. According to [16] most high energy electrons will be shielded by a layer thicker than 300 mils (7.6 mm) of Al, being most of the dose caused, behind shielding, by high energy protons. On the other hand the dose rate caused by high energy protons from the Van Allen Belts reduces approximately 10 times per inch of Al equivalent shielding. The shielding done by the satellite mass of 22 kg distributed around the sensor would explain the low dose rate measured on the board. Unfortunately, proprietary information of the satellite manufacturer could not be disclosed to us to make a more complete analysis based on Monte Carlo simulations.

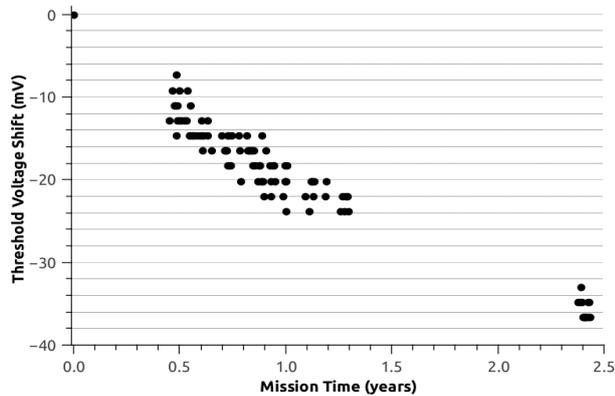

Fig. 6. Threshold Voltage Shift as a function of time for the dosimeter on the MeMOSat board.

## V. Conclusions and future work

The work presents the design and realization of MOS dosimetry on the MeMOSat board, mounted on the BugSat-1 satellite, on a 620 km low earth orbit. The sensors used in this work where COTS MOS transistors, proving the feasibility of using such devices in low cost experiments. The dose measured on the board after 2.5 years is ~1.2 krad, a value compatible with the fact that the board was protected by the mass of the rest of the satellite from most of the high energy particles in the space environment.

The work is being continued, and the MeMOSat board has been replaced by a more complex experiment named LabOSat [20]. During the past years, the FOXFET, a new thick gate oxide MOS dosimeter was developed and tested in different application [9]. The FOXFET proved to have lower fading and a much higher sensitivity, and will replace the COTS sensors used in this work. However, low cost academic CubeSat projects might use COTS MOS transistors as an effective way to estimate dose and perform experiments in orbit. The use of different sensors in different positions of the satellite to evaluate the relative dose rates observed after different shielding thicknesses. Also the long term fading needs to be taken in account for a reliable determination of the real dose in the experiment.